\documentclass[english,aps,prb,twocolumn]{revtex4}
\usepackage{amsmath}
\usepackage{amssymb}
\usepackage{amsfonts}
\usepackage{amsthm}
\usepackage{graphicx}
\providecommand{\tabularnewline}{\\}

\begin{document}

\title{Evaluation of the nondiabaticity of quantum
molecular dynamics with the dephasing representation of quantum fidelity}

\author{Tom\'{a}\v{s} Zimmermann}

\author{Ji\v{r}\'{\i} Van\'{\i}\v{c}ek}

\email{jiri.vanicek@epfl.ch}

\affiliation{Laboratory of Theoretical Physical Chemistry, Institut des Sciences
et Ing\'{e}nierie Chimiques, Ecole Polytechnique F\'{e}d\'{e}rale de Lausanne
(EPFL), CH-1015, Lausanne, Switzerland}

\date{\today}

\begin{abstract}
We propose an approximate method for evaluating the importance of 
non-Born-Oppenheimer effects on the quantum dynamics of nuclei.
The method uses a generalization of the dephasing representation (DR)
of quantum fidelity to several diabatic potential energy surfaces
and its computational cost is the cost of dynamics
of a classical phase space distribution. It can be implemented easily 
into any molecular dynamics program and also can utilize on-the-fly \textit{ab initio} electronic structure information. 
 We test the methodology on
three model problems introduced by Tully and on the photodissociation of NaI. The results show that for dynamics close to the diabatic limit
the decay of fidelity due to nondiabatic effects is described accurately
by the DR. In this regime, unlike the mixed quantum-classical methods such as surface hopping or Ehrenfest dynamics, the DR can capture
more subtle quantum effects than the population transfer between potential energy
surfaces. Hence we propose using the DR to estimate the \emph{dynamical} importance
of diabatic, spin-orbit, or other couplings between potential energy surfaces. The acquired information
can help reduce the complexity of a studied system without affecting
the accuracy of the quantum simulation. 
\end{abstract}
\maketitle

\section{Introduction\label{sec:Introduction}}

The nonadiabatic effects play an important role in many chemical phenomena
and often must be taken into account in accurate calculations of molecular
properties such as spectra or reaction rates.\cite{Butler1998,Worth2004,FDisscuss2004}
Many nonadiabatic quantum (QM) simulations are performed in the diabatic
(or quasi-diabatic) basis, which offers several computational advantages,
especially in the vicinity of conical intersections of adiabatic potential
energy surfaces (PESs). Usually, more PESs must be included to describe
a system of interest accurately. One may ask: Which diabatic surfaces are important? How much is the result affected
by neglecting the less important surfaces? The method we propose below
quantifies the importance of couplings between PESs and thus can
help in answering these questions.

A direct way to determine whether the coupling of diabatic PESs affects
the result of a simulation is to compute the desired quantity by running
QM dynamics for both the uncoupled and coupled systems, and to
compare the results. Here, we propose a more general approach, which can give the information 
about the importance of couplings for all observables at the same time. It consists
in comparing the wave functions by computing the QM
fidelity, defined as\cite{peres:1984}
\begin{equation}
F_{\mathrm{QM}}(t)=\left|f(t)\right|^{2}=\left|\left\langle \psi_{0}(t)|\psi_{p}(t)\right\rangle \right|^{2}.
\label{eq:qm_fidelity}
\end{equation}
In the general setting, $\psi_{0}(t)$ and $\psi_{p}(t)$ are wave
functions evolved in the unperturbed and perturbed systems, respectively.
In our case the {}``unperturbed'' represents the \emph{uncoupled}
system, {}``perturbed'' the \emph{coupled} system, and $\psi$ denotes
the full molecular wave function, i.e., it includes both nuclear and
electronic degrees of freedom. Values of $F_{\mathrm{QM}}(t)$ close
to unity imply that the perturbation is not important: the uncoupled
Hamiltonian can be used in quantitative simulations. Values significantly
below unity imply that the perturbation is important: the affected
PESs and couplings should be included in the simulation.

Instead of computing fidelity directly from the 
definition \eqref{eq:qm_fidelity}, one can use the dephasing
representation (DR),\cite{Vanicek2004,Vanicek2004a,Vanicek2006} which is an efficient semiclassical approximation of fidelity.\cite{Cerruti2002,Vanicek2003} Recently, the
DR was used successfully to evaluate the accuracy of QM dynamics
on a single but approximate PES.\cite{Li2009,Zimmermann2010}
Below, we generalize this method to several surfaces and use it to
evaluate how the dynamics is affected by couplings between the diabatic PESs. 

Unlike the computational cost of a direct QM calculation,
the cost of the DR does not
grow exponentially with the number $D$ of degrees of freedom.\cite{Vanicek2006,Li2009} Hence the DR can be used for many-dimensional systems inaccessible to current methods of QM dynamics. An advantage of the DR in comparison
with mixed QM-classical methods for nonadiabatic dynamics, such as the Ehrenfest dynamics, various surface hopping methods,\cite{Tully1971,Tully1990,Nielsen2000,Heller2002} or methods in which the classical limit is obtained by the linearization
of the QM propagator,\cite{Dunkel2008} is that the DR, being
a semiclassical method, takes the nuclear coherence effects
into account at least approximately. Other semiclassical approaches to nonadiabatic dynamics
exist, including, amongst others, the multiple-spawning methods\cite{Martinez1996,Yang2009}
or methods\cite{Miller2009} based on the Miller-Meyer-Stock-Thoss
classical electron model. The advantage of the DR is that, unlike
the majority of semiclassical approaches, the DR does not require
the Hessian of the potential energy, which is the most expensive element
of first-principles semiclassical dynamics methods (see, e.g., Ref.
\onlinecite{Ceotto2009a}). 

Clearly, the above mentioned advantages do not come for free:
First, unlike most other approaches noted above, the DR is not a general dynamical method. It can only describe properties expressible in terms of fidelity. Second, the DR in the diabatic basis is expected to work accurately mainly when
the coupling is relatively weak or the nuclear velocities are large, i.e., when the dynamics is close
to the diabatic limit. Nevertheless, in many cases the DR works very well even for strong
perturbations that lead to completely different classical phase space
distributions for uncoupled and coupled Hamiltonians.

Applications for which the DR method is particularly well suited include
the chemical reactions which proceed near the diabatic limit. The
photodissociation of bromopropionyl chloride or bromoacetyl chloride\cite{Waschewsky1994,Valero2006}
serve as examples. Another possible application would be to quantify
the relative importance of Hamiltonian coupling terms of various origins, e.g., of the spin-orbit vs. diabatic couplings. For instance, in the Cl($^{\text{2}}$P)
+ H$_{\text{2}}$ reaction the spin-orbit coupling terms clearly dominate
over diabatic couplings, which can therefore be neglected in a simulation.
\cite{Alexander2004}

\section{Theory\label{sec:Theory}}

The starting point is the Hamiltonian describing nuclear motion in
a molecule, expressed in the diabatic basis. To compute the decay
of fidelity due to the coupling between $n$ PESs, the Hamiltonian
$n\times n$ matrix is split into the uncoupled (i.e., diagonal) part
$\mathbf{\hat{H}}^{\mathrm{diag}}$ and the coupling (i.e., offdiagonal) part
$\Delta\mathbf{\hat{V}}$, 
\begin{eqnarray}
\mathbf{\hat{H}}&=&\mathbf{\hat{H}}^{\mathrm{diag}}+\Delta\mathbf{\hat{V}},\label{eq:Hamiltonian} \\
\mathbf{\hat{H}}^{\mathrm{diag}}&=&\mathbf{\hat{T}}+\mathbf{\hat{V}}^{\mathrm{diag}}\quad\mathrm{and}\quad\Delta\mathbf{\hat{V}}=\mathbf{\hat{V}}^{\mathrm{offdiag}}.\label{eq:Hamiltonian_parts}
\end{eqnarray}
 In Eq. \eqref{eq:Hamiltonian_parts}, $\mathbf{\hat{T}}$ is the
diagonal nuclear kinetic energy matrix and $\mathbf{\hat{V}}^{\mathrm{diag}}$
contains the diabatic PESs, which are uncoupled
in $\mathbf{\hat{H}}^{\mathrm{diag}}$ and coupled in $\mathbf{\hat{H}}$ by the
elements of $\mathbf{\hat{V}}^{\mathrm{offdiag}}$. ({\bf Bold} face denotes $n \times n$ matrices, hat $\hat{}$ denotes operators.)

To derive the DR, one starts from the expression for QM
fidelity amplitude, applicable to both pure and mixed states,\cite{Vanicek2006} and generalized to the multi-PES setting,
\begin{equation}
f_{\mathrm{QM}}(t)=\mathrm{Tr}\left(e^{-i\mathbf{\hat{H}}^{\mathrm{diag}}t/\hbar}\mathbf{\mathbf{\boldsymbol{\hat{\rho}}}}\,e^{+i\mathbf{\hat{H}}t/\hbar}\right),\label{eq:f_rho}
\end{equation}
where $\boldsymbol{\hat{\rho}}$ is the density operator of the initial
state. Generalizing the derivation from Ref. \onlinecite{Vanicek2010}, fidelity amplitude can be written exactly as
\begin{equation}
f_{\mathrm{QM}}(t)=\mathrm{Tr}\int dx\mathbf{\mathbf{\boldsymbol{\rho}}_{\mathrm{W}}}(x)\cdot \left(e^{+i\mathbf{\hat{H}}t/\hbar} e^{-i\mathbf{\hat{H}}^{\mathrm{diag}}t/\hbar}\right)_{\mathrm{W}}\!\!(x),
\end{equation}
where $\boldsymbol{\rho}_{\mathrm{W}}$ is the Wigner transform
of the initial state $\boldsymbol{\hat{\rho}}$,
\[
\left[\rho_{\mathrm{W}}\right]_{ij}(x)=h^{-D}\!\!\int d\xi\left\langle q-\frac{\xi}{2}\right|\boldsymbol{\hat{\rho}}\left|q+\frac{\xi}{2}\right\rangle \exp\left(i\frac{\xi\cdot p}{\hbar}\right),
\]
and $x$ denotes the point $(q,p)$ in the $2 \times D$-dimensional phase space.
Approximating the Wigner transform of the product of the time-evolution operators, one arrives at the DR expression 
\begin{align}
f_{\mathrm{DR}}(t) &=\mathrm{Tr}\left[ \int dx^{0}\mathbf{\mathbf{\boldsymbol{\rho}}_{\mathrm{W}}}(x^{0})\cdot e^{-i \Delta\mathbf{S}\left(x^{0},t\right)/\hbar}\right] ,\label{eq:f_DR} \\
\Delta\mathbf{S}\left(x^{0},t\right) &=\int_{0}^{t}d\tau\Delta\mathbf{V}_{\mathrm{W}}\left[x^{\tau}\left(x^{0}\right)\right],\label{eq:delta_S}
\end{align}
where $\Delta\mathbf{S}\left(x^{0},t\right)$
is the action at time $t$ due to Wigner representation $\Delta\mathbf{V_{\mathrm{W}}}$
of $\Delta\mathbf{\hat{V}}$ along the trajectory $x^{\tau}$ of $\mathbf{H}^{\mathrm{diag}}$.
Note that $\Delta\mathbf{V_{\mathrm{W}}}$, $\Delta\mathbf{S}$, and $\mathbf{\boldsymbol{\rho}}_{\mathrm{W}}$
are matrix quantities. If $\Delta\mathbf{\hat{V}}$ contains only
diabatic coupling elements, then $\Delta\mathbf{\hat{V}}\equiv\Delta\mathbf{V}(\hat{q})$
and $\Delta\mathbf{V_{\mathrm{W}}}(x)=\Delta\mathbf{V}(q)$. For initial
Gaussian wave packets, the Wigner function equals the classical phase
space density which is strictly a probability distribution.

Since $\mathrm{Tr}\int dx^{0}\mathbf{\mathbf{\boldsymbol{\rho}}_{\mathrm{W}}}(x^{0})=1$,
it follows from Eq. \eqref{eq:f_DR} that $f_{\mathrm{DR}}$ can be
computed as a Monte Carlo average $\left\langle \exp(-i\Delta\mathbf{S}(x^{0},t)/\hbar\right\rangle _{\boldsymbol{\rho}_{\mathrm{W}}(x^{0})}$
with initial conditions sampled from the Wigner distribution $\boldsymbol{\rho}_{\mathrm{W}}(x^{0})$
of the initial state.  However, as $\Delta\mathbf{S}$ in Eq.~(\ref{eq:delta_S}) is much smaller than the action in other SC methods, the DR alleviates the notorious ``sign problem.''

If the initial state lies on a single surface
$n$, i.e., if $\left[\rho_{\mathrm{W}}\right]_{nj}=0$ and $\left[\rho_{\mathrm{W}}\right]_{jj}=0$
for $j\ne n$, then no other elements of $\mathbf{V}=\mathbf{V}^{\mathrm{diag}}+\mathbf{V}^{\mathrm{offdiag}}$
than $V_{nn}$ and $V_{nj},j\neq n$, enter the calculation. This means
that no information about other diagonal elements $V_{jj}$ is used
in the DR calculation. Thus, except for special cases, $F_{\mathrm{DR}}$
is expected to approximate $F_{\mathrm{QM}}$ accurately only when
the detailed structure of the remaining PESs does not significantly
affect the dynamics on $V_{nn}$.

\section{Results and discussion \label{sec:Results}}

\begin{figure}[hptb]
\includegraphics[width=1\columnwidth]{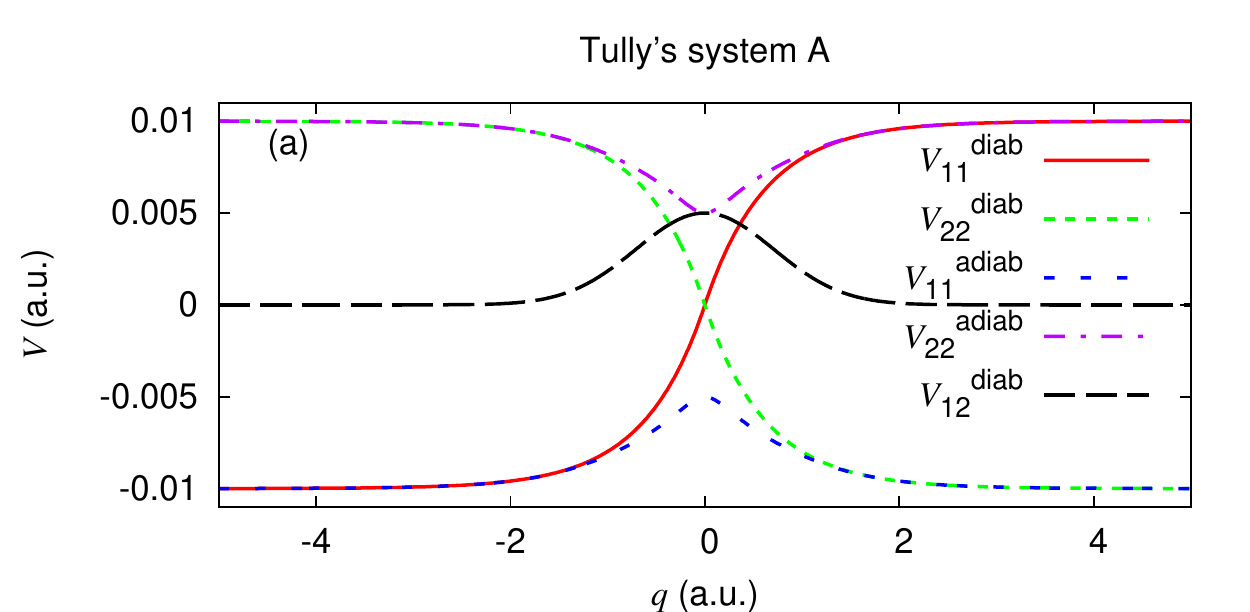}\tabularnewline
\includegraphics[width=1\columnwidth]{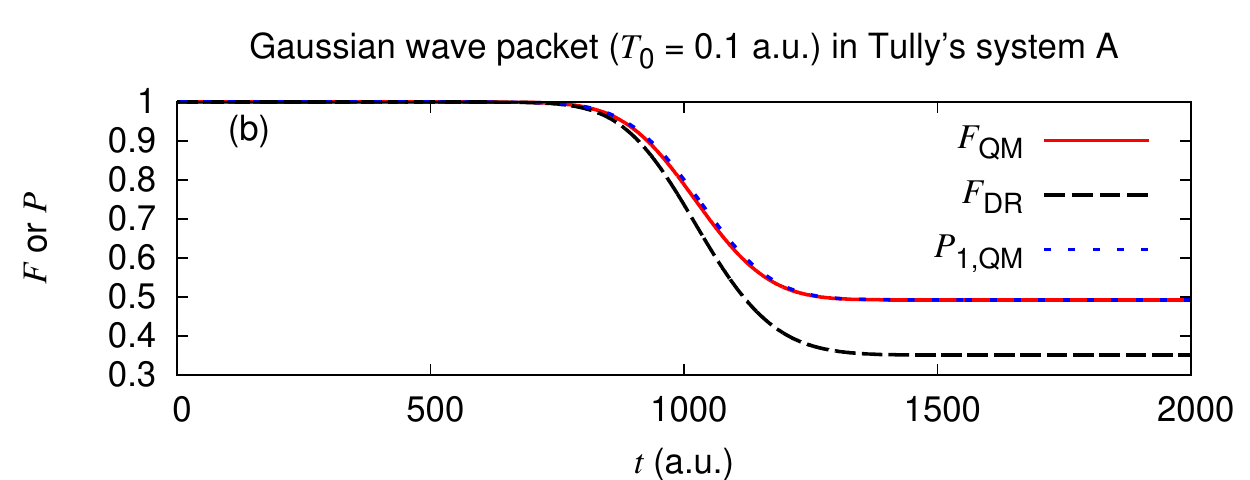}\tabularnewline
\includegraphics[width=1\columnwidth]{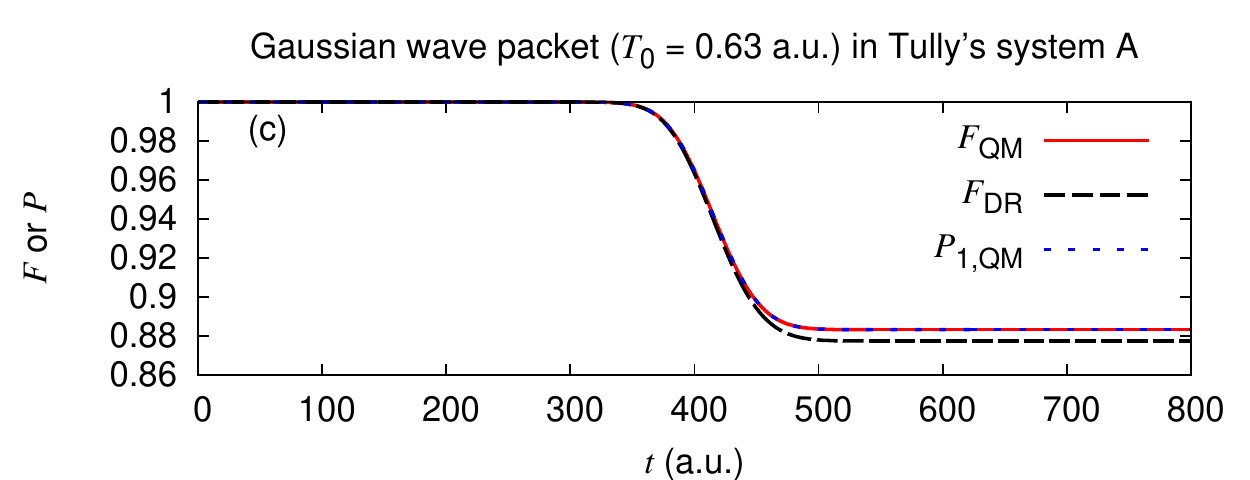}\tabularnewline
\caption{Fidelity in Tully's problem A. (a) The diabatic and adiabatic PESs
and the diabatic coupling. (b) and (c): Quantum fidelity $F_{\mathrm{QM}}$,
the quantum survival probability $P_{1,\mathrm{QM}}$ on the initial
PES, and $F_{\mathrm{DR}}$ as functions of time for two different
values of initial kinetic energy $T_0$. \label{fig:Tully_A}}
\end{figure}

\begin{figure}[hptb]
\includegraphics[width=1\columnwidth]{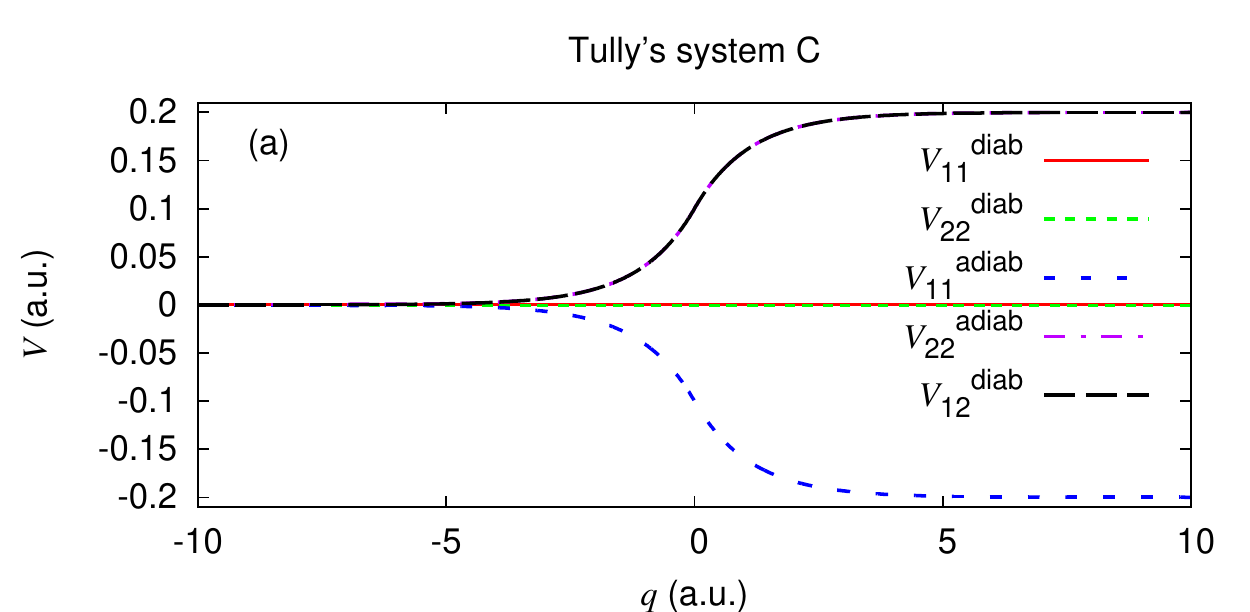}\tabularnewline
\includegraphics[width=1\columnwidth]{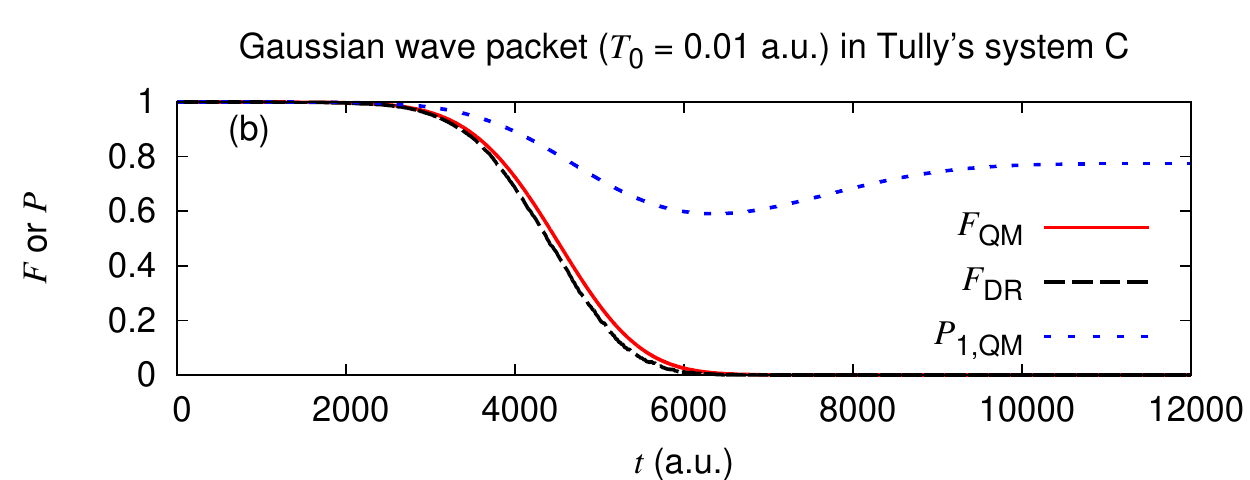}\tabularnewline
\includegraphics[width=1\columnwidth]{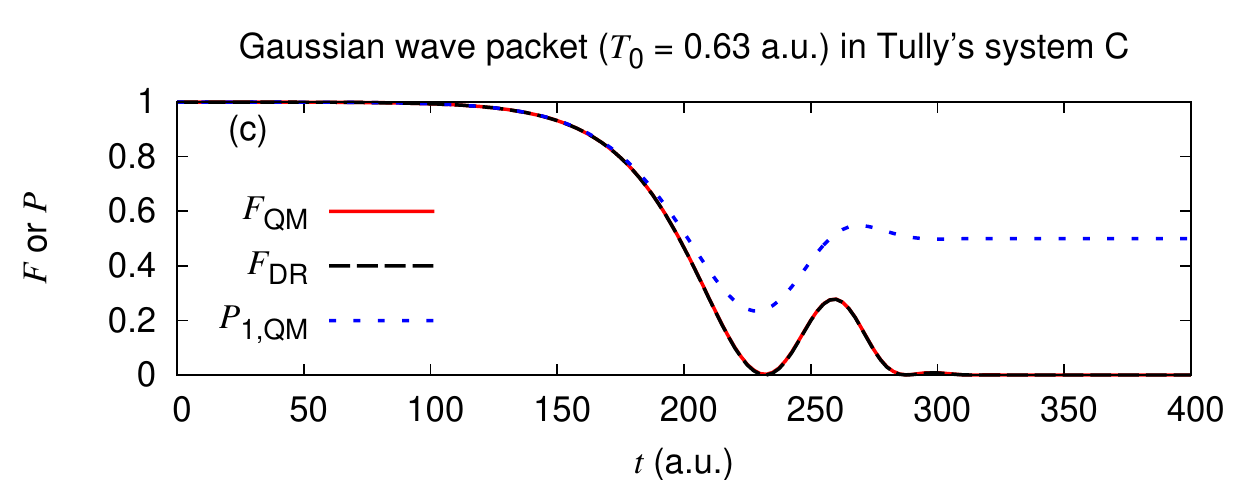}\tabularnewline
\caption{Fidelity in Tully's problem C.\label{fig:Tully_C}}
\end{figure}

\textit{Tully's model problems.} First, the method is tested on three
one-dimensional model potentials proposed by Tully\cite{Tully1990}
to cover the most important characteristics of nonadiabatic transitions.
Diabatic and adiabatic PESs as well as the coupling
terms $V_{12}$ for Tully's problems A and C are shown in Figs. \ref{fig:Tully_A}-\ref{fig:Tully_C};
further details can be found in Ref. \onlinecite{Tully1990}.

In all cases, the initial wave packet was a Gaussian with approximately
10 \% dispersion in momentum and located on the lower (problems A and B) or upper (problem C) energy diabatic
PES in the asymptotic region without diabatic couplings. The equations
of motion were integrated until the wave packet left the interaction
region. 

The simple avoided crossing model (problem A in the original paper\cite{Tully1990})
represents the most often encountered situation. As can be seen in
Fig. \ref{fig:Tully_A}, the survival probability $P_{1,\mathrm{QM}}=\mathrm{Tr}\boldsymbol{\hat{\rho}}_{11}(t)$
on the initial PES is nearly equal to the QM fidelity, suggesting
that the fidelity decay is caused almost exclusively by the transition
to the second diabatic PES. 
Close to the diabatic limit, $F_{\mathrm{DR}}$
accurately reproduces $F_{\mathrm{QM}}$. In practice, the error of
$F_{\mathrm{DR}}$ comes from the intrinsic error of the approximation
and from the statistical error due to the finite number of trajectories.
In this case, the intrinsic relative error is $\sim$ 0.5 \% when
fidelity decays by 10 \% to 0.9 and even less for the wave packets
with higher initial momentum and hence fidelity. The relative statistical
error remains below 1\% even when only a single trajectory is used,
since most of the fidelity decay is due to the transitions to the
second surface. For slower wave packets, fidelity decreases, and the
agreement between $F_{\mathrm{QM}}$ and $F_{\mathrm{DR}}$ stays
qualitative, with $F_{\mathrm{DR}}$ always decaying faster than $F_{\mathrm{QM}}$.
Finally, for a fixed initial momentum, decreasing the coupling $V_{12}$
results in a slower decay of fidelity but the relative error of $(F_{\mathrm{QM}}-F_{\mathrm{DR}})/(1-F_{\mathrm{QM}})$
is approximately independent of the coupling. Conversely, if $V_{12}$ is increased substantially, fidelity initially decays quickly to zero, which is captured well by the DR. Subsequently, fidelity may rise again, which is usually not well reflected by the DR.

In the dual avoided crossing model (problem B, not shown), the DR
works the best again for high energy wave packets. For those
the detailed structure of the second PES does not significantly affect
the motion on the first PES. At low energies, where a significant
transfer of probability density to the diabatic surface $V_{22}$
and back occurs, the DR fails to reproduce the QM
fidelity even qualitatively. Nevertheless, in this case both $F_{\mathrm{DR}}$
and $F_{\mathrm{QM}}$ initially decay almost to zero (although they
might rise again later), correctly reflecting that the coupling is
important and should not be neglected.

A very interesting situation occurs in the extended coupling model
(problem C), where the two diabatic PESs $V_{11}$ and $V_{22}$ are
almost equal (they differ just by a small constant shift) but the
adiabatic surfaces are well separated due to the coupling.
At very low energies of the wave packet {[}Fig. \ref{fig:Tully_C}
(b)], fidelity $F$ goes to zero despite that the survival probability $P_1$
on the upper surface remains close to unity. Therefore, in contrast
to previously discussed cases, it is not possible to estimate the
extent of the nondiabaticity of QM dynamics on the basis of electronic transitions
only. 
Unlike survival probability, fidelity describes the nondiabaticity
effects on the coherent nuclear dynamics correctly. It is remarkable that the DR describes $F$ so accurately despite that the DR dynamics on the diabatic surface ignores the reflection of a large part of the QM wavepacket from the upper adiabatic surface.
The decrease of fidelity is correctly captured by $F_{\mathrm{DR}}$
for all energies of the wavepacket. At higher energies [Fig. \ref{fig:Tully_C}
(c)], fidelity converges to zero after several oscillations, whereas
the survival probability converges to $1/2$. At very high energies
(not shown), two wave packets moving on the two PESs interfere for
a long time, $F_{\mathrm{DR}}$ and $F_{\mathrm{QM}}$ agree and oscillate
between zero and unity. Surface hopping and Ehrenfest dynamics might predict $P_1$ quite well, but since $V_{11} \approx V_{22}$, surface hopping would incorrectly predict that $F \approx P_1$. While Ehrenfest dynamics would predict a decay of $F$ below $P$, it is unlikely that it would capture $F$ quantitatively since the QM wave packet splits into a faster and slower components whereas Ehrenfest dynamics uses a single mean-field surface.
One of the reasons why the DR performs so
well in the extended coupling model is the similarity of the diabatic
PESs in the coupling region. If the surfaces are different,
e.g., when $V_{22}=cx$, the fidelity
decay at low energies is still well approximated. However, at higher
energies, $F_{\mathrm{DR}}$ oscillations slowly dephase from $F_{\mathrm{QM}}$,
since the DR neglects effects of $V_{22}$.

\begin{figure}[hptb]
\includegraphics[width=1\columnwidth]{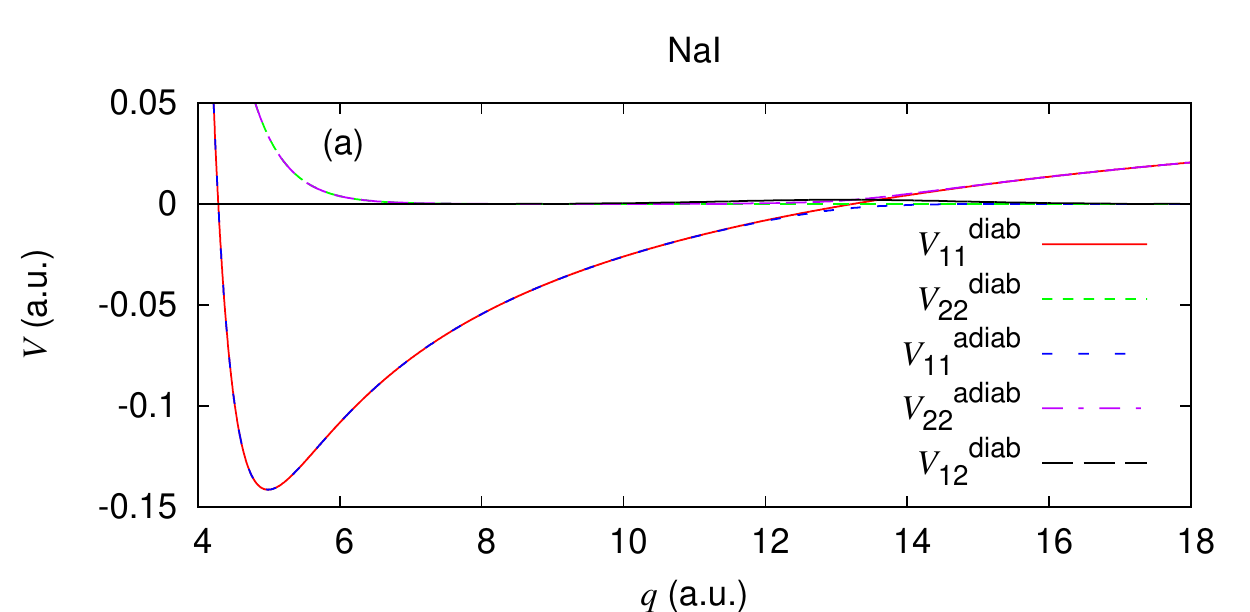}\tabularnewline
\includegraphics[width=1\columnwidth]{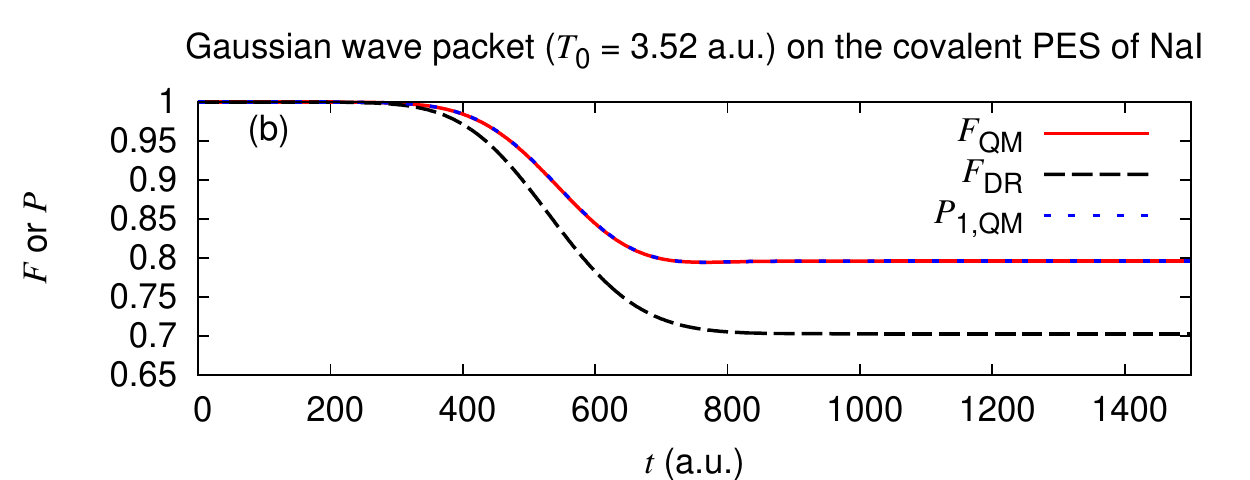}\tabularnewline
\includegraphics[width=1\columnwidth]{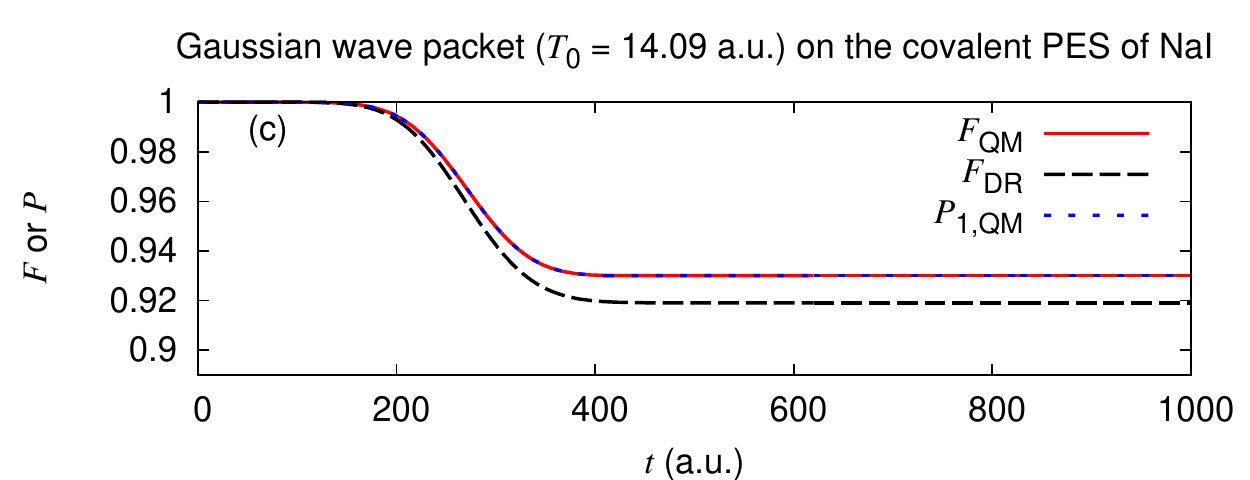}\tabularnewline
\caption{Fidelity in the photodissociation of NaI.\label{fig:NaI}}
\end{figure}

\textit{Photodissociation of NaI.} We also applied the methodology
to the photodissociation of NaI using a two-surface model of Engel
and Metiu\cite{Faist1976,Veen1981,Engel1989} (see Fig. \ref{fig:NaI}).
In the original experiment of Mokhtari \textit{et al.}\cite{Mokhtari1990}
the molecule was excited by the light in the 310-390 $\mathrm{nm}$
range, which led to only a weakly nonadiabatic motion of the wave
packet on the excited surface. In this regime (corresponding to $T_0\approx0.07\:\mathrm{a.u.}$
at the plateau of the PES), far from the diabatic limit, $F_{\mathrm{DR}}$
does not quantitatively reproduce $F_{\mathrm{QM}}$. However, when
the momentum is approximately 10 times higher and the dissociation
of NaI almost diabatic, $F_{\mathrm{DR}}$ reproduces $F_{\mathrm{QM}}$
rather well.

\textit{Computational details.} QM dynamics calculations used
the first-order split operator method. Classical trajectories were
computed using the first-order symplectic Euler algorithm. The time
step varied from 0.02 $\mathrm{a.u.}$ to 1 $\mathrm{a.u.}$, depending
on the system and initial momentum. 

To guarantee convergence, 16384 classical paths were used to produce
the DR plots. The statistical error of $f_{\mathrm{DR}}$ for a two-surface 
system due to a finite number $N$ of paths is given by
\begin{equation}
\sigma_{\mathrm{stat}}^{2}=\frac{1}{N}(\left\langle \cos^{2}\left(\varphi\right)\right\rangle -\left\langle \cos\left(\varphi\right)\right\rangle ^{2}),\label{eq:stat_error}
\end{equation}
where $\varphi=\Delta\mathbf{S}_{12}\left(x^{0},t\right)/\hbar$ is the phase accumulated along a trajectory 
and $\left\langle \cos\left(\varphi\right)\right\rangle=f_{\mathrm{DR}}$.
As demonstrated on Tully's model A, where a single trajectory was
sufficient, a much lower number of trajectories than 16384 is needed
to obtain an accurate estimate of fidelity in cases where fidelity
stays close to unity. Equation (\ref{eq:stat_error}) implies that for a given value of $\sigma_{\mathrm{stat}}$ and $f_{\mathrm{DR}}$, the number $N$ of trajectories needed is approximately independent of dimensionality.

\textit{Conclusions}. Presented results demonstrate the utility of
the DR in analyzing the molecular QM dynamics involving multiple
PESs. On one hand, in the nearly diabatic regime, $F_{\mathrm{DR}}$
accurately approximates $F_{\mathrm{QM}}$. On the other hand, in
systems far from the diabatic limit $F_{\mathrm{DR}}$ decays quickly
to zero and thus detects the importance of nondiabatic couplings although
it may not reproduce $F_{\mathrm{QM}}$ accurately. Hence, the method
can be used to establish the level of nondiabaticity of QM dynamics,
without the need for a QM dynamics simulation. In fact, we propose the condition $F \approx 1$ (instead of the standard requirement of high survival probability) as the rigorous definition of the diabatic limit (Fig.~\ref{fig:Tully_C}). Nevertheless, for single avoided crossings, e.g.,  fidelity could be used to estimate the survival probability and hence the branching ratios (Figs.~\ref{fig:Tully_A} and \ref{fig:NaI}). 

The DR calculation can be performed easily for all systems accessible
to classical molecular dynamics and for which coupling elements $V_{ij},i\neq j$,
are available. However, it remains to be verified how the method will
perform in higher-dimensional systems and in systems with more than
two important surfaces. The first issue was partially addressed in Ref.
\onlinecite{Li2009} where the DR was applied to the two-dimensional
photodissociation of CO$_{\text{2}}$, confirming that neither
the accuracy nor the number of classical trajectories needed are significantly
affected by increased dimensionality.

At the moment we are exploring a related and complementary problem
to that of the present paper, namely whether the DR in the adiabatic
basis can estimate the level of nonadiabaticity of QM dynamics
near the adiabatic limit.

This research was supported by the Swiss NSF (Grant No. $200021\_124936/1$)
and by the EPFL. We thank Ivano Tavernelli for helpful discussions.

\bibliographystyle{apsrev} 


\begin{thebibliography}{28}
\expandafter\ifx\csname natexlab\endcsname\relax\def\natexlab#1{#1}\fi
\expandafter\ifx\csname bibnamefont\endcsname\relax
  \def\bibnamefont#1{#1}\fi
\expandafter\ifx\csname bibfnamefont\endcsname\relax
  \def\bibfnamefont#1{#1}\fi
\expandafter\ifx\csname citenamefont\endcsname\relax
  \def\citenamefont#1{#1}\fi
\expandafter\ifx\csname url\endcsname\relax
  \def\url#1{\texttt{#1}}\fi
\expandafter\ifx\csname urlprefix\endcsname\relax\def\urlprefix{URL }\fi
\providecommand{\bibinfo}[2]{#2}
\providecommand{\eprint}[2][]{\url{#2}}

\bibitem[{\citenamefont{Butler}(1998)}]{Butler1998}
\bibinfo{author}{\bibfnamefont{L.~J.} \bibnamefont{Butler}},
  \bibinfo{journal}{Annu. Rev. Phys. Chem.} \textbf{\bibinfo{volume}{49}},
  \bibinfo{pages}{125} (\bibinfo{year}{1998}).

\bibitem[{\citenamefont{Worth and Cederbaum}(2004)}]{Worth2004}
\bibinfo{author}{\bibfnamefont{G.~A.} \bibnamefont{Worth}} \bibnamefont{and}
  \bibinfo{author}{\bibfnamefont{L.~S.} \bibnamefont{Cederbaum}},
  \bibinfo{journal}{Annu. Rev. Phys. Chem.} \textbf{\bibinfo{volume}{55}},
  \bibinfo{pages}{127} (\bibinfo{year}{2004}).

\bibitem[{\citenamefont{Child and Robb}(2004)}]{FDisscuss2004}
\bibinfo{editor}{\bibfnamefont{M.~S.} \bibnamefont{Child}} \bibnamefont{and}
  \bibinfo{editor}{\bibfnamefont{M.~A.} \bibnamefont{Robb}}, eds.,
  \emph{\bibinfo{title}{Non-Adiabatic Effects in Chemical Dynamics}}, vol.
  \bibinfo{volume}{127} of \emph{\bibinfo{series}{Faraday Discussions}}
  (\bibinfo{publisher}{Royal Society fo Chemistry}, \bibinfo{year}{2004}).

\bibitem[{\citenamefont{Peres}(1984)}]{peres:1984}
\bibinfo{author}{\bibfnamefont{A.}~\bibnamefont{Peres}},
  \bibinfo{journal}{Phys. Rev. A} \textbf{\bibinfo{volume}{{\bf 30}}},
  \bibinfo{pages}{1610} (\bibinfo{year}{1984}).

\bibitem[{\citenamefont{Van\'{i}\v{c}ek}(2004{\natexlab{a}})}]{Vanicek2004}
\bibinfo{author}{\bibfnamefont{J.}~\bibnamefont{Van\'{i}\v{c}ek}},
  \bibinfo{journal}{Phys. Rev. E} \textbf{\bibinfo{volume}{70}},
  \bibinfo{pages}{055201} (\bibinfo{year}{2004}{\natexlab{a}}).

\bibitem[{\citenamefont{Van\'{i}\v{c}ek}(2004{\natexlab{b}})}]{Vanicek2004a}
\bibinfo{author}{\bibfnamefont{J.}~\bibnamefont{Van\'{i}\v{c}ek}}
  (\bibinfo{year}{2004}{\natexlab{b}}), \eprint{arXiv:quant-ph/0410205v1}.

\bibitem[{\citenamefont{Van\'{i}\v{c}ek}(2006)}]{Vanicek2006}
\bibinfo{author}{\bibfnamefont{J.}~\bibnamefont{Van\'{i}\v{c}ek}},
  \bibinfo{journal}{Phys. Rev. E} \textbf{\bibinfo{volume}{73}},
  \bibinfo{pages}{046204} (\bibinfo{year}{2006}).

\bibitem[{\citenamefont{Cerruti and Tomsovic}(2002)}]{Cerruti2002}
\bibinfo{author}{\bibfnamefont{N.~R.} \bibnamefont{Cerruti}} \bibnamefont{and}
  \bibinfo{author}{\bibfnamefont{S.}~\bibnamefont{Tomsovic}},
  \bibinfo{journal}{Phys. Rev. Lett.} \textbf{\bibinfo{volume}{{\bf 88}}},
  \bibinfo{pages}{054103} (\bibinfo{year}{2002}).

\bibitem[{\citenamefont{Van\'{i}\v{c}ek and Heller}(2003)}]{Vanicek2003}
\bibinfo{author}{\bibfnamefont{J.}~\bibnamefont{Van\'{i}\v{c}ek}}
  \bibnamefont{and} \bibinfo{author}{\bibfnamefont{E.~J.}
  \bibnamefont{Heller}}, \bibinfo{journal}{Phys. Rev. E}
  \textbf{\bibinfo{volume}{68}}, \bibinfo{pages}{056208}
  (\bibinfo{year}{2003}).

\bibitem[{\citenamefont{Li et~al.}(2009)\citenamefont{Li, Mollica, and
  Van\'{i}\v{c}ek}}]{Li2009}
\bibinfo{author}{\bibfnamefont{B.}~\bibnamefont{Li}},
  \bibinfo{author}{\bibfnamefont{C.}~\bibnamefont{Mollica}}, \bibnamefont{and}
  \bibinfo{author}{\bibfnamefont{J.}~\bibnamefont{Van\'{i}\v{c}ek}},
  \bibinfo{journal}{J. Chem. Phys.} \textbf{\bibinfo{volume}{131}},
  \bibinfo{eid}{041101} (\bibinfo{year}{2009}).

\bibitem[{\citenamefont{Zimmermann et~al.}(2010)\citenamefont{Zimmermann,
  Ruppen, Li, and Van\'{i}\v{c}ek}}]{Zimmermann2010}
\bibinfo{author}{\bibfnamefont{T.}~\bibnamefont{Zimmermann}},
  \bibinfo{author}{\bibfnamefont{J.}~\bibnamefont{Ruppen}},
  \bibinfo{author}{\bibfnamefont{B.}~\bibnamefont{Li}}, \bibnamefont{and}
  \bibinfo{author}{\bibfnamefont{J.}~\bibnamefont{Van\'{i}\v{c}ek}},
  \bibinfo{journal}{Int. J. Quantum Chem.}  (\bibinfo{year}{2010}).

\bibitem[{\citenamefont{Tully and Preston}(1971)}]{Tully1971}
\bibinfo{author}{\bibfnamefont{J.~C.} \bibnamefont{Tully}} \bibnamefont{and}
  \bibinfo{author}{\bibfnamefont{R.~K.} \bibnamefont{Preston}},
  \bibinfo{journal}{J. Chem. Phys.} \textbf{\bibinfo{volume}{55}},
  \bibinfo{pages}{562} (\bibinfo{year}{1971}).

\bibitem[{\citenamefont{Tully}(1990)}]{Tully1990}
\bibinfo{author}{\bibfnamefont{J.~C.} \bibnamefont{Tully}},
  \bibinfo{journal}{J. Chem. Phys.} \textbf{\bibinfo{volume}{93}},
  \bibinfo{pages}{1061} (\bibinfo{year}{1990}).

\bibitem[{\citenamefont{Nielsen et~al.}(2000)\citenamefont{Nielsen, Kapral, and
  Ciccotti}}]{Nielsen2000}
\bibinfo{author}{\bibfnamefont{S.}~\bibnamefont{Nielsen}},
  \bibinfo{author}{\bibfnamefont{R.}~\bibnamefont{Kapral}}, \bibnamefont{and}
  \bibinfo{author}{\bibfnamefont{G.}~\bibnamefont{Ciccotti}},
  \bibinfo{journal}{J. Stat. Phys.} \textbf{\bibinfo{volume}{101}},
  \bibinfo{pages}{225} (\bibinfo{year}{2000}).

\bibitem[{\citenamefont{Heller et~al.}(2002)\citenamefont{Heller, Segev, and
  Sergeev}}]{Heller2002}
\bibinfo{author}{\bibfnamefont{E.~J.} \bibnamefont{Heller}},
  \bibinfo{author}{\bibfnamefont{B.}~\bibnamefont{Segev}}, \bibnamefont{and}
  \bibinfo{author}{\bibfnamefont{A.~V.} \bibnamefont{Sergeev}},
  \bibinfo{journal}{J. Phys. Chem. B} \textbf{\bibinfo{volume}{106}},
  \bibinfo{pages}{8471} (\bibinfo{year}{2002}).

\bibitem[{\citenamefont{Dunkel et~al.}(2008)\citenamefont{Dunkel, Bonella, and
  Coker}}]{Dunkel2008}
\bibinfo{author}{\bibfnamefont{E.~R.} \bibnamefont{Dunkel}},
  \bibinfo{author}{\bibfnamefont{S.}~\bibnamefont{Bonella}}, \bibnamefont{and}
  \bibinfo{author}{\bibfnamefont{D.~F.} \bibnamefont{Coker}},
  \bibinfo{journal}{J. Chem. Phys.} \textbf{\bibinfo{volume}{129}},
  \bibinfo{eid}{114106} (\bibinfo{year}{2008}).

\bibitem[{\citenamefont{Mart\'{i}nez et~al.}(1996)\citenamefont{Mart\'{i}nez,
  Ben-Nun, and Levine}}]{Martinez1996}
\bibinfo{author}{\bibfnamefont{T.~J.} \bibnamefont{Mart\'{i}nez}},
  \bibinfo{author}{\bibfnamefont{M.}~\bibnamefont{Ben-Nun}}, \bibnamefont{and}
  \bibinfo{author}{\bibfnamefont{R.~D.} \bibnamefont{Levine}},
  \bibinfo{journal}{J. Phys. Chem.} \textbf{\bibinfo{volume}{100}},
  \bibinfo{pages}{7884} (\bibinfo{year}{1996}).

\bibitem[{\citenamefont{Yang et~al.}(2009)\citenamefont{Yang, Coe, Kaduk, and
  Mart\'{i}nez}}]{Yang2009}
\bibinfo{author}{\bibfnamefont{S.}~\bibnamefont{Yang}},
  \bibinfo{author}{\bibfnamefont{J.~D.} \bibnamefont{Coe}},
  \bibinfo{author}{\bibfnamefont{B.}~\bibnamefont{Kaduk}}, \bibnamefont{and}
  \bibinfo{author}{\bibfnamefont{T.~J.} \bibnamefont{Mart\'{i}nez}},
  \bibinfo{journal}{J. Chem. Phys.} \textbf{\bibinfo{volume}{130}},
  \bibinfo{pages}{134113} (\bibinfo{year}{2009}).

\bibitem[{\citenamefont{Miller}(2009)}]{Miller2009}
\bibinfo{author}{\bibfnamefont{W.~H.} \bibnamefont{Miller}},
  \bibinfo{journal}{J. Phys. Chem. A} \textbf{\bibinfo{volume}{113}},
  \bibinfo{pages}{1405} (\bibinfo{year}{2009}).

\bibitem[{\citenamefont{Ceotto et~al.}(2009)\citenamefont{Ceotto, Atahan,
  Tantardini, and Aspuru-Guzik}}]{Ceotto2009a}
\bibinfo{author}{\bibfnamefont{M.}~\bibnamefont{Ceotto}},
  \bibinfo{author}{\bibfnamefont{S.}~\bibnamefont{Atahan}},
  \bibinfo{author}{\bibfnamefont{G.~F.} \bibnamefont{Tantardini}},
  \bibnamefont{and}
  \bibinfo{author}{\bibfnamefont{A.}~\bibnamefont{Aspuru-Guzik}},
  \bibinfo{journal}{J. Chem. Phys.} \textbf{\bibinfo{volume}{130}},
  \bibinfo{eid}{234113} (\bibinfo{year}{2009}).

\bibitem[{\citenamefont{Waschewsky et~al.}(1994)\citenamefont{Waschewsky, Kash,
  Myers, Kitchen, and Butler}}]{Waschewsky1994}
\bibinfo{author}{\bibfnamefont{G.~C.~G.} \bibnamefont{Waschewsky}},
  \bibinfo{author}{\bibfnamefont{P.~W.} \bibnamefont{Kash}},
  \bibinfo{author}{\bibfnamefont{T.~L.} \bibnamefont{Myers}},
  \bibinfo{author}{\bibfnamefont{D.~C.} \bibnamefont{Kitchen}},
  \bibnamefont{and} \bibinfo{author}{\bibfnamefont{L.~J.}
  \bibnamefont{Butler}}, \bibinfo{journal}{J. Chem. Soc., Faraday Trans.}
  \textbf{\bibinfo{volume}{90}}, \bibinfo{pages}{1581} (\bibinfo{year}{1994}).

\bibitem[{\citenamefont{Valero and Truhlar}(2006)}]{Valero2006}
\bibinfo{author}{\bibfnamefont{R.}~\bibnamefont{Valero}} \bibnamefont{and}
  \bibinfo{author}{\bibfnamefont{D.~G.} \bibnamefont{Truhlar}},
  \bibinfo{journal}{J. Chem. Phys.} \textbf{\bibinfo{volume}{125}},
  \bibinfo{eid}{194305} (\bibinfo{year}{2006}).

\bibitem[{\citenamefont{Alexander et~al.}(2004)\citenamefont{Alexander,
  Capecchi, and Werner}}]{Alexander2004}
\bibinfo{author}{\bibfnamefont{M.~H.} \bibnamefont{Alexander}},
  \bibinfo{author}{\bibfnamefont{G.}~\bibnamefont{Capecchi}}, \bibnamefont{and}
  \bibinfo{author}{\bibfnamefont{H.~J.} \bibnamefont{Werner}},
  \bibinfo{journal}{Faraday Discuss.} \textbf{\bibinfo{volume}{127}},
  \bibinfo{pages}{59} (\bibinfo{year}{2004}).

\bibitem[{\citenamefont{Van\'{i}\v{c}ek
  et~al.}(2010)\citenamefont{Van\'{i}\v{c}ek, Mollica, Prosen, and
  Strunz}}]{Vanicek2010}
\bibinfo{author}{\bibfnamefont{J.}~\bibnamefont{Van\'{i}\v{c}ek}},
  \bibinfo{author}{\bibfnamefont{C.}~\bibnamefont{Mollica}},
  \bibinfo{author}{\bibfnamefont{T.}~\bibnamefont{Prosen}}, \bibnamefont{and}
  \bibinfo{author}{\bibfnamefont{W.}~\bibnamefont{Strunz}},
  \bibinfo{journal}{to be published}  (\bibinfo{year}{2010}).

\bibitem[{\citenamefont{Faist and Levine}(1976)}]{Faist1976}
\bibinfo{author}{\bibfnamefont{M.~B.} \bibnamefont{Faist}} \bibnamefont{and}
  \bibinfo{author}{\bibfnamefont{R.~D.} \bibnamefont{Levine}},
  \bibinfo{journal}{J. Chem. Phys.} \textbf{\bibinfo{volume}{64}},
  \bibinfo{pages}{2953} (\bibinfo{year}{1976}).

\bibitem[{\citenamefont{van Veen et~al.}(1981)\citenamefont{van Veen, Vries,
  Sokol, Baller, and de~Vries}}]{Veen1981}
\bibinfo{author}{\bibfnamefont{N.}~\bibnamefont{van Veen}},
  \bibinfo{author}{\bibfnamefont{M.~D.} \bibnamefont{Vries}},
  \bibinfo{author}{\bibfnamefont{J.}~\bibnamefont{Sokol}},
  \bibinfo{author}{\bibfnamefont{T.}~\bibnamefont{Baller}}, \bibnamefont{and}
  \bibinfo{author}{\bibfnamefont{A.}~\bibnamefont{de~Vries}},
  \bibinfo{journal}{Chem. Phys.} \textbf{\bibinfo{volume}{56}},
  \bibinfo{pages}{81 } (\bibinfo{year}{1981}).

\bibitem[{\citenamefont{Engel and Metiu}(1989)}]{Engel1989}
\bibinfo{author}{\bibfnamefont{V.}~\bibnamefont{Engel}} \bibnamefont{and}
  \bibinfo{author}{\bibfnamefont{H.}~\bibnamefont{Metiu}}, \bibinfo{journal}{J.
  Chem. Phys.} \textbf{\bibinfo{volume}{90}}, \bibinfo{pages}{6116}
  (\bibinfo{year}{1989}).

\bibitem[{\citenamefont{Mokhtari et~al.}(1990)\citenamefont{Mokhtari, Cong,
  Herek, and Zewail}}]{Mokhtari1990}
\bibinfo{author}{\bibfnamefont{A.}~\bibnamefont{Mokhtari}},
  \bibinfo{author}{\bibfnamefont{P.}~\bibnamefont{Cong}},
  \bibinfo{author}{\bibfnamefont{J.~L.} \bibnamefont{Herek}}, \bibnamefont{and}
  \bibinfo{author}{\bibfnamefont{A.~H.} \bibnamefont{Zewail}},
  \bibinfo{journal}{Nature} \textbf{\bibinfo{volume}{348}},
  \bibinfo{pages}{225} (\bibinfo{year}{1990}).

\end{thebibliography}
\end{document}